\documentclass[a4paper,fleqn,usenatbib]{mnras}

\usepackage{newtxtext,newtxmath}
\usepackage{times}

\usepackage[T1]{fontenc}
\usepackage{ae,aecompl}

\usepackage{graphicx}	
\usepackage{amsmath}	
\usepackage{amssymb}

\newcommand{\cgsflux}{erg\,s$^{-1}$\,cm$^{-2}$}
\newcommand{\cgslum}{erg\,s$^{-1}$}
\newcommand{\p}{$\phantom{+}$}

\title[Candidate novae in early-type galaxies]
{Discovering novae in early-type galaxies with MUSE: 
A chance find in NGC\,1404, and twelve more candidates from an archival search}

\author[Russell J. Smith]{
	Russell J. Smith\thanks{E-mail: russell.smith@durham.ac.uk}\\
	Centre for Extragalactic Astronomy, University of Durham, Durham DH1 3LE, United Kingdom}

\date{Submitted to MNRAS Letters, 22nd January 2020}

\pubyear{2020}

\begin{document}
\label{firstpage}
\pagerange{\pageref{firstpage}--\pageref{lastpage}}
\maketitle

\begin{abstract}
I report the discovery of a transient broad-H$\alpha$ point source in the outskirts of the giant elliptical galaxy NGC\,1404, discovered in archival observations taken with the MUSE integral field spectrograph.  The H$\alpha$ line width of 1950\,km\,s$^{-1}$ FWHM, and luminosity of (4.1$\pm$0.1)$\times${}$10^{36}$\,\cgslum, are consistent with a nova outburst, and the source is not visible in MUSE data obtained nine months later. A transient soft X-ray source was detected at the same position (within $<$1\,arcsec), 14 years before the H$\alpha$ transient. If the X-ray and H$\alpha$ emission are from the same object, the source may be a short-timescale recurrent nova with a massive white dwarf accretor, and hence a possible Type-Ia supernova progenitor. Selecting broad-H$\alpha$ point sources in MUSE archival observations for a set of nearby  early-type galaxies, I discovered twelve more nova candidates with similar properties to the NGC\,1404 source, including five in NGC\,1380 and four in NGC\,4365. Multi-epoch data are available for four of these twelve sources; all four are confirmed to be transient on $\sim$1\,year timescales, supporting their identification as novae.
\end{abstract}

\begin{keywords}
galaxies: elliptical and lenticular, cD -- novae, cataclysmic variables 
\end{keywords}

\section{Introduction}

Nova eruptions are caused by thermonuclear detonation of material accreted onto the surface of a white dwarf (WD) from a companion star \citep[see][for extensive reviews]{2008clno.book.....B}. The outburst expels part of the accreted envelope, but does not destroy the system, so it is thought that novae are inherently recurrent, though relatively few have undergone multiple {\it recorded} eruptions \citep[$\sim$30 in the Milky Way, M31, and the Large Magellanic Cloud; see e.g.][]{2019arXiv190910497D}.
If these episodes lead to net growth in the WD mass \citep{2016ApJ...819..168H}, then novae are important as a class of potential progenitors for Type-Ia supernovae.

Novae were historically considered promising as extragalactic distance indicators, leading to several systematic searches in galaxies beyond the Local Group, including early-type targets \citep{2002Sci...296.1275D,2003ApJ...599.1302F,2005ApJ...618..692N,2015ApJ...811...34C,2016ApJS..227....1S}.
While the classical decline-rate versus peak luminosity relation no 
longer seems  viable as a distance indicator
\citep[e.g.][]{2017ApJ...839..109S}, 
the {\it rate} of nova production in early-type galaxies is also of interest, as a probe of close binary stars in stellar populations different than those in the Milky Way. Luminosity-normalised nova rates of $\nu_K$\,$\sim$\,(1--3)\,yr$^{-1}$\,$(10^{10}L_{\odot,K})^{-1}$ are typical \citep[e.g.][]{2003ApJ...599.1302F,2015ApJ...811...34C}.
Population synthesis models suggest that $\nu_K$ should be larger for younger populations \citep{2016MNRAS.458.2916C}, but observed differences between spirals and ellipticals appear to be modest, and the 
bulges of M\,31 and M\,81 seem to exhibit larger $\nu_K$ than their 
disks \citep{1987ApJ...318..520C,2004AJ....127..816N,2006MNRAS.369..257D}.

\begin{figure*}
\begin{center}
\vskip 0mm
\includegraphics[width=108mm]{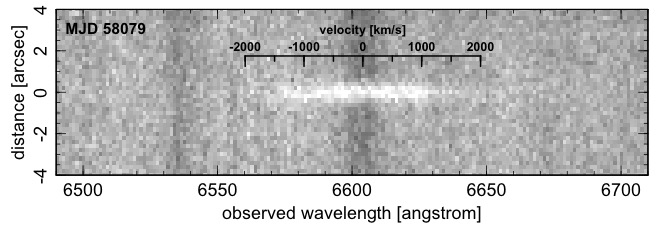} 
\includegraphics[width=60mm]{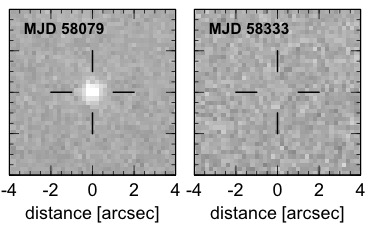}
\end{center}
\vskip -7mm
\caption{Left: Two-dimensional spectrum of NGC1404-S1 from MUSE, in a similar format to that used in discovering the source. Effectively this is a spectrum from a pseudo-slit of width 1\,arcsec centred on the source, 37\,arcsec from the galaxy centre.
The narrow H$\alpha$ absorption line from the nearly uniform 
diffuse light of the galaxy is seen as the dark vertical band, 
while the much broader point source emission appears as the bright horizontal streak.
Right: Continuum-subtracted H$\alpha$ images of the source in the discovery data, and in a second epoch observation 254 days later.}
\label{fig:twodspec}
\end{figure*}

\begin{figure*}
\begin{center}
\vskip -6mm 
\includegraphics[width=175mm]{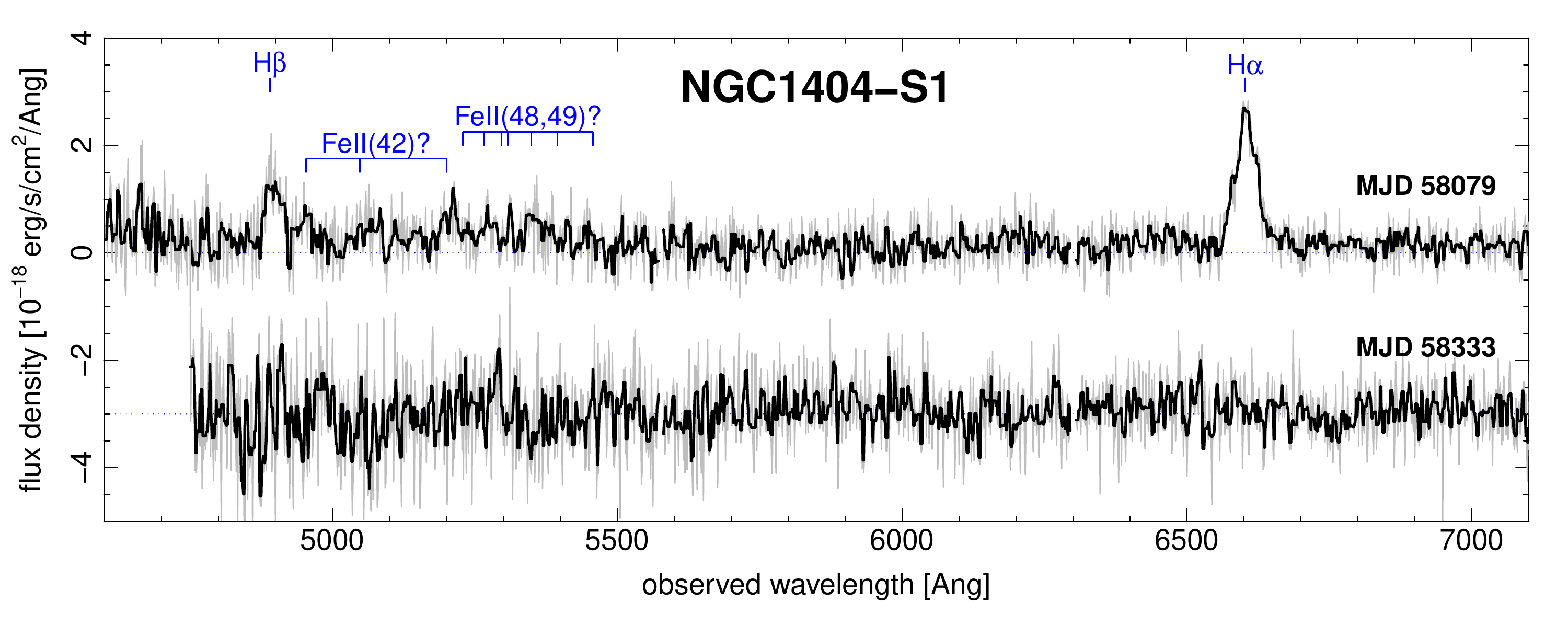}
\end{center}
\vskip -6mm
\caption{Extracted spectrum of NGC1404-S1, after subtracting the diffuse background, showing the H$\alpha$ and H$\beta$ emission lines in the discovery epoch spectrum (above) and the absence of emission in the later epoch (below, offset by 3$\times$10$^{-18}$\,\cgsflux). The grey trace shows the original spectrum, while the black trace includes a 5-pixel (6.25\,\AA) median smoothing. No lines are detected in NGC1404-S1 beyond the spectral range shown.}
\label{fig:onedspec}
\end{figure*}

In the initial period after maximum light, the H$\alpha$ emission from a nova fades much more slowly than the optical continuum \citep*[][]{1983ApJ...272...92C}. The longer visibility time in H$\alpha$ has long been exploited in narrow-band photometric searches for extra-galactic novae \citep[e.g.][]{1990AJ.....99.1079C}, typically using fixed narrow-band filters of $\sim$75\,\AA\ width. 
In general, wide-field integral-field spectroscopy (IFS) provides a more flexible approach to detecting emission-line sources, by matching the wavelength range to the 
width of the features being sought. Moreover, IFS data provides the full spectroscopic coverage simultaneously with the search process, which is not available from fixed- or tunable-filter observations. 
The only example of this approach so far is by \citet*{2018MNRAS.473.4130M}, who detected five M\,31 novae in H$\alpha$, using single-epoch imaging fourier-transform spectrometer observations.

In this {\it Letter} I describe how novae can be identified at distances of $\sim$20\,Mpc, using the more orthodox Multi-Unit Spectroscopic Explorer (MUSE) IFS on the ESO Very Large Telescope \citep{2010SPIE.7735E..08B}.
The work was prompted by the chance discovery of a transient H$\alpha$ source in NGC\,1404, with properties consistent with a nova, as summarized in Section~\ref{sec:n1404s1}. Section~\ref{sec:x34} notes a spatially coincident transient X-ray source, detected more than a decade earlier than the H$\alpha$ detection, suggesting this nova is a recurrent
  system. Motivated by the NGC\,1404 discovery, Section~\ref{sec:more} describes a search for further novae in archival MUSE  observations of nearby early-type galaxies, from which twelve more candidates are presented. Section~\ref{sec:concs} summarizes the work, highlighting the feasibility of a dedicated extragalactic nova survey with MUSE.

\section{A transient broad H$\alpha$  source in NGC\,1404}\label{sec:n1404s1}

The source described in this section, hereafter NGC1404-S1, was discovered serendipitously during visual inspection of MUSE data for NGC\,1404, a massive elliptical galaxy in the Fornax Cluster.
The observation was obtained on 2017 Nov 22 (MJD 58079) for programme 296.B-5054, `The Fornax3D Survey' \citep{2018A&A...616A.121S}. The total exposure time is 3600\,sec, the point spread function FHWM is $\sim$0.7\,arcsec, and the data
were obtained on a photometrically stable night.

Viewing the archived data-cube in `pseudo-slit' format, i.e. with one spatial and one spectral dimension, and scanning across the second spatial direction, revealed a striking broad emission feature centred on the much narrower stellar H$\alpha$ absorption line, as reproduced in Figure~\ref{fig:twodspec}.
The emission line is spatially unresolved, and projected 37\,arcsec (3.6\,kpc) from centre of NGC\,1404, approximately twice the half-light radius \citep{2003AJ....125..525J}.
After extracting a one-dimensional spectrum, and subtracting the local background, a corresponding H$\beta$ line is also visible, confirming the source as being associated with NGC\,1404, rather than an active galactic nucleus in the background  (Figure~\ref{fig:onedspec}). The H$\alpha$ luminosity is (4.1$\pm$0.1)$\times${}$10^{36}$\,\cgslum, while the line width is 1950\,km\,s$^{-1}$ FWHM. 
The position of NGC1404-S1 is covered by another MUSE observation, obtained 
on 2018 Aug 03 (MJD 58333). 
NGC1404-S1 is undetected in this second-epoch observation, with continuum and line emission consistent with zero.

The most plausible interpretation of NGC1404-S1 is as a nova.
The luminosity 
is consistent with the characteristic peak output of novae \citep[10$^{36-37}$\,\cgslum\ at maximum;][]{1990ApJ...356..472C}, 
and line-widths of $\sim$2000\,km\,s$^{-1}$ are also typical \citep[e.g.][]{2011ApJ...734...12S}. 
Given the H$\alpha$ luminosity, the typical 
continuum-to-line flux ratios 
for novae $\ga$15\,days after maximum \cite[converted from  $B$$-$H$\alpha$\,$\approx$\,2.7 in the system of][]{1990AJ.....99.1079C} correspond to an expected $B$-band continuum flux of $\sim$3$\times$10$^{-19}$\,\cgsflux\,\AA$^{-1}$, which is compatible with the measured signal at the blue end of the MUSE spectrum (2.6$\pm$0.05\,$\times$10$^{-19}$\,\cgsflux\,\AA$^{-1}$).
The disappearance of the source after 254\,days is expected for 
typical nova H$\alpha$ decline rates.
The Balmer decrement in novae is commonly steeper than for the standard Case-B recombination ratios, with H$\alpha$/H$\beta$\,=\,5--10. The H$\beta$ line is clearly detected in NGC1404-S1, however, with a ratio of $\sim$3, 
indicating little deviation from the Case-B value of 2.7.
Novae are classified into Fe\,{\sc ii} and He/N spectral types, dependent on the strongest non-Balmer lines present, with the majority in the former class \citep{1992AJ....104..725W}. The NGC1404-S1 spectrum cannot be classified confidently, but there is a hint that Fe\,{\sc ii}\,(42) could be present at low level, and  blended Fe\,{\sc ii}\,(48,49) lines might contribute to the apparent broad flux excess at 5200--5400\,\AA.

The luminosity of NGC\,1404, integrated over the region sampled by the discovery image, is 1.0\,$\times$\,$10^{11}$\,$L_{\odot,K}$ 
\citep[measured from 2 Micron All-Sky Survey Large Galaxy Atlas images;][]{2003AJ....125..525J}.
Hence, for 
the typical specific nova rate of 
$\nu_K$\,$\sim$\,2\,yr$^{-1}$\,$(10^{10}L_{\odot,K})^{-1}$, 
there are $\sim$20\,eruptions per year.
For a typical H$\alpha$ decay rate of 0.03\,mag\,day$^{-1}$ \citep{1990AJ.....99.1079C},
and assuming S1 was detected close to maximum, and could be confidently detected at 25 per cent of its observed flux, then the visibility period is $\sim$0.14\,yr. Thus the mean number of detectable eruptions in the frame should be $\sim$3. Even allowing for the greater challenge of detection against the brightest parts of the continuum light, observing an ongoing nova in this frame (or indeed two novae, see Section~\ref{sec:more}) is not implausible.

\begin{figure*}
\begin{center}
\vskip 0mm
\includegraphics[width=180mm]{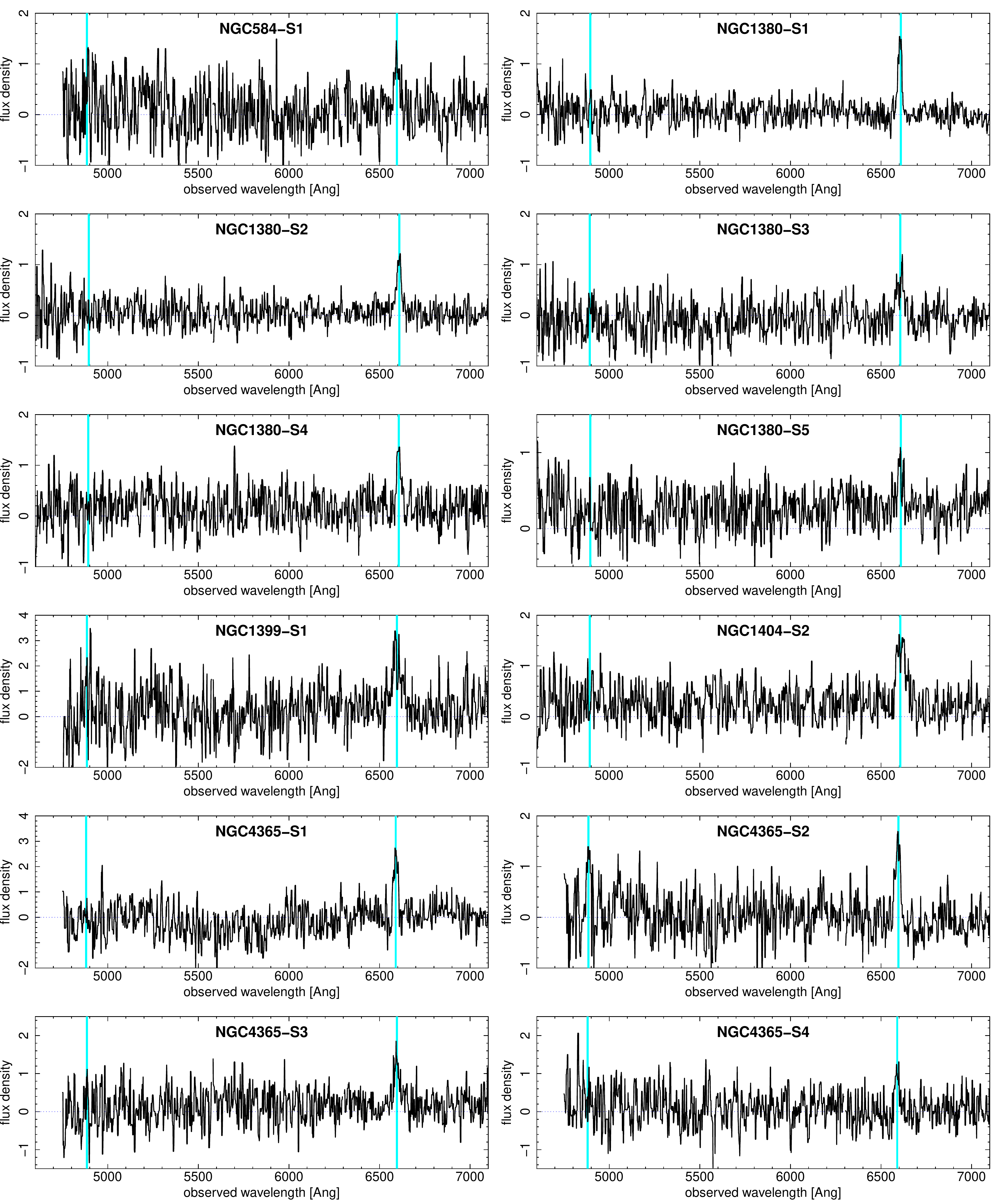}
\end{center}
\vskip -2mm
\caption{Additional nova candidates selected as broad H$\alpha$ point sources in nearby early-type galaxies.
The flux density unit is $10^{-18}$\,\cgsflux\,\AA$^{-1}$.
For clarity only the 5-pixel smoothed spectra are shown. Cyan lines mark H$\alpha$ and the 
expected location of H$\beta$ (which is generally not detected).}
\label{fig:faintspectra}
\end{figure*}


\begin{table*}
    \centering
        \caption{Candidate novae detected in MUSE observations of nearby early-type galaxies. MJD is the Modified Julian Date for the discovery observation epoch, or the time difference from discovery  for other epochs. 
        The H$\alpha$ FWHM is determined from a simple gaussian fit to the discovery epoch spectrum.
        The line fluxes are measured within an aperture of diameter 1.2\,arcsec, and $\pm$1.06\,FWHM, i.e. $\pm$2.5\,$\sigma$. The adopted distances are
        from  \citet{2001ApJ...546..681T} for NGC\,584, and
      \citet{2009ApJ...694..556B} for the other galaxies.
        The candidates from Section~\ref{sec:more} are ordered by host-galaxy R.A. and H$\alpha$ significance.
        }
    \begin{tabular}{lcccccccc}
    \hline
        Host & distance  & Source & R.A. & Dec  &  H$\alpha$ FWHM  & MJD &  Flux  & Luminosity \\        
                &  [Mpc] &  & \multicolumn{2}{c}{[J2000]} & [km\,s$^{-1}$] & [days] & [10$^{-17}$\,\cgsflux] & [10$^{36}$\,\cgslum] \\        
        \hline

         NGC\,1404 & 20.2 & S1 & 03:38:51.54 & --35:35:01.5  & 1947 &  58079 & \p8.40$\pm$0.23 & \p4.10$\pm$0.11\\ 
           & &  &  &  &  & \,\,+254 & \p0.17$\pm$0.43 & \p0.08$\pm$0.21\\
        \hline
        NGC\,584 & 20.1 & S1 & 01:31:18.28  & --06:52:37.0  & 1433 &  57570 & \p2.87$\pm$0.45 & \p1.39$\pm$0.22  \\
        
        NGC\,1380 & 21.2 &  S1 & 03:36:29.61 &  --34:57:35.2 & 1023 & 57773 & \p3.12$\pm$0.18 & \p1.68$\pm$0.10 \\ 
                           & &  &  &  &  & \,\,+294 & --0.07$\pm$0.18 & --0.04$\pm$0.10\\
        & & S2 & 03:36:29.47 & --34:57:57.2 & 1568 & 57773 & \p2.90$\pm$0.25 & \p1.56$\pm$0.13 \\ 
                   & &  &  &  &  & \,\,+294 & --0.74$\pm$0.34 & --0.40$\pm$0.18\\

         & & S3 & 03:36:28.55 & --34:58:57.0  & 2110 & 57753  & \p3.37$\pm$0.41 & \p1.81$\pm$0.22\\
        & & S4 &  03:36:28.47 &  --34:58:07.2 & 1199 & 57753 & \p2.56$\pm$0.35 & \p1.38$\pm$0.19 \\
                   & &  &  &  &  & \,\,\,\,\,+20 & \p1.47$\pm$0.35 & \p0.79$\pm$0.19\\
                   & &  &  &  &  & \,\,+314 & \p0.32$\pm$0.37 & \p0.17$\pm$0.20\\
        & & S5 & 03:36:29.47 & --34:57:25.3 & 1486 & 58067 & \p1.22$\pm$0.24 & \p0.66$\pm$0.13 \\ 
                 & &  &  &  &  & \,\,--294 & --0.23$\pm$0.20 & --0.13$\pm$0.11\\
        NGC\,1399 & 20.9 & S1 & 03:38:27.75 & --35:27:11.7 & 3693 & 56944 & \p8.28$\pm$1.34 & \p4.33$\pm$0.77 \\ 
                 NGC\,1404 & 20.2 & S2 & 03:38:51.08 & --35:35:17.8 &  2971 & 58079 & \p6.18$\pm$0.52 & \p3.02$\pm$0.25 \\

        NGC\,4365 & 23.1 & S1 & 12:24:28.01 & +07:18:49.1 & 1199 & 57066 & \p5.52$\pm$0.49 & \p3.52$\pm$0.31 \\ 
         &  & S2 & 12:24:28.50 & +07:19:33.5 & 1659 &  57066 & \p4.31$\pm$0.41 & \p2.77$\pm$0.26 \\ 
        & & S3 & 12:24:29.41 & +07:19:23.9 & 1804 & 57066 & \p3.89$\pm$0.48 & \p2.48$\pm$0.31 \\ 
        & & S4 & 12:24:26.91 & +07:18:53.5 & 1101 & 57066 & \p2.60$\pm$0.42 & \p1.66$\pm$0.27  \\ 
                 \hline
    \end{tabular}
    \label{tab:sources}
\end{table*}

\section{X-rays and relationship to NGC1404-X34}\label{sec:x34}

Novae are associated with `super-soft' X-ray emission from the burning WD, which follows the optical outburst by tens to hundreds of days, when the ejected material becomes optically thin to $\sim$0.5\,keV photons. 
A search for X-ray emission from NGC1404-S1 yielded a $<$1\,arcsec match to NGC1404-X34 in the {\it Chandra X-ray Observatory} source catalogue of \cite{2016ApJS..224...40W}. NGC1404-X34 was detected
with a 0.3--8.0\,keV flux of 4.5$\times${}$10^{15}$\,\cgsflux, corresponding to luminosity 2.2$\times${}$10^{38}$\,\cgslum\ in a 45\,ksec observation taken on 2003 May 28 (MJD 52788). However, it is absent from
a 29\,ksec observation on 2003 Feb 03 (105\,days earlier), and from all 
other {\it Chandra} observations of NGC\,1404. 
From its hardness ratios, the 2003 detection is categorised as a `quasi-soft source', in the scheme of \cite{2003astro.ph.11374D}.

Since the optical emission from novae precedes the X-rays, the 2003 {\it Chandra} detection does not relate to the 2017 outburst, but would 
have to be associated with a previous outburst, occurring 14 years before to the MUSE observation.
The X-ray luminosity of NGC1404-X34 is close to the maximum expected in theoretical models of post-nova sources \citep{2016MNRAS.455..668S}. Sources with such high luminosity require WDs close to the Chandrasekhar mass, and have higher effective temperatures ($kT$\,$\approx$\,100\,eV) 
than lower-luminosity systems, which would be in line with the 
`quasi-soft' rather than `super-soft' designation for NGC1404-X34.
Eruptions on the most massive WDs are also expected to have the shortest recurrence timescales ($\la$1\,year) and the briefest period of X-ray emission ($\la$100\,days)  \citep{2005ApJ...623..398Y}.
Such a short X-ray lifetime would be compatible with the appearance within three months in 2003, while frequent recurrence would be consistent with the NGC1404-X34 in 2003 and S1 in 2017 being different (and not necessarily consecutive) eruptions of the same system. Such high-mass WD systems are predicted to be rare in ellipticals, however \citep{2016MNRAS.458.2916C}, and it may seem overly fortuitous to have observed the short-lived X-ray outburst.

The X-ray properties of NGC1404-X34/S1 bear interesting comparison with nova M31N 2008-12a, which has the shortest known recurrence time (1\,year), and is also thought to be a massive WD accretor \citep{2014A&A...563L...9D,2014A&A...563L...8H,2014ApJ...786...61T}, as well as one of the strongest candidates for a single-degenerate type Ia supernova progenitor.

\section{An archival search for similar sources}\label{sec:more}

Motivated by the discovery of NGC1404-S1, I have made a preliminary search for sources with similar properties in the MUSE archive. 
The goal here is not to generate a statistically robust sample, but simply to determine whether similar objects can be {\it routinely} detected in MUSE observations of nearby galaxies.

I searched for broad-H$\alpha$ point sources in MUSE data for twelve galaxies in total  
(NGC\,584,
NGC\,1332,
NGC\,1380,
NGC\,1399,
NGC\,1404, 
NGC\,1407,
NGC\,1428, 
NGC\,3311,
NGC\,4374, 
NGC\,4365,
NGC\,4473, 
and NGC\,5419),
selected to be nearby early-type galaxies
having observations longer than 1800\,sec in the ESO archive as of December 2019, but excluding those 
with complex nebular H$\alpha$ emission (e.g. M\,87).
Overlapping observations separated by more than one night were treated separately. I made use of additional pointings in the outskirts of galaxies where available (e.g. NGC\,1380 and NGC\,1404), to enlarge the search area.
From each datacube, I constructed net-H$\alpha$ emission-line images with bandwidths of 25 and 60\,\AA\ (1100 and 2700 km\,s$^{-1}$), applied a 2.5\,arcsec FWHM unsharp-mask filter to suppress continuum-subtraction mismatch and any diffuse emission, and searched visually for spatially compact residuals. The spectrum of each candidate was then inspected to confirm that the line is broad. (A single bright narrow-line candidate was rejected, being likely a H\,{\sc ii} region or dwarf galaxy.)

The archived MUSE data are nominally calibrated to a physical flux scale by the observatory pipeline. Most of the observations 
for the detected nova candidates were obtained during nights with stable photometric conditions (assessed from the seeing-monitor flux variation in the ESO database), and the flux calibration for these was adopted at face value. For one observation made in poorer conditions (NGC\,4365), the photometric zero point was checked independently against the $r$-band aperture flux of the galaxy from Sloan Digital Sky Survey imaging.

Table~\ref{tab:sources} and Figure~\ref{fig:faintspectra} show  twelve additional nova candidate selected by this process, including a second source in NGC\,1404, five in NGC\,1380 (also in Fornax) and four in NGC\,4365 (in the background of Virgo). In seven targets, no novae were detected. The range in detection number should not be taken to reflect a true difference in nova rate, given the small numbers and the absence of any corrections for variation in the extent and quality of the data.

The selected sources are characterized by the absence of bright lines other than H$\alpha$, and little if any continuum emission.
Apart from H$\alpha$, the only convincing line detections are for H$\beta$ in NGC4365-S2 and for O\,{\sc i} 8446\,\AA\ (beyond the spectral range shown in the figure) for NGC1380-S1. 
Most of the sources are fainter than NGC1404-S1, with 
luminosities in the range (0.6--4)$\times$10$^{36}$\,\cgslum. 
The line widths are generally 1000--3000\,km\,s$^{-1}$ FWHM, consistent with the bandwidths used in the selection process.

Multi-epoch data are available for five of the sources selected for broad H$\alpha$ emission (including NGC1404-S1). In {\it all} of these cases, the line is absent in exposures taken $>$200\,days before or after from the detection epoch; hence this selection method seems in general to yield transient sources. Disappearance on this timescale is consistent with the H$\alpha$ fading rates of
of 0.01--0.08\,mag\,day$^{-1}$ for  M\,31 novae \cite{1990ApJ...356..472C}.
In the case of NGC1380-S4, three epochs are available. The H$\alpha$ line fades by a factor of 1.75$\pm$0.50 in the 20 days between the first and second epoch, i.e. 0.030$\pm$0.015\,mag\,day$^{-1}$.
The third epoch is 294 days later and the emission is no longer detectable.

None of the additional sources has an X-ray counterpart in the \cite{2016ApJS..224...40W} catalogue.

\section{Summary and outlook}\label{sec:concs}

The goal of this {\it Letter} has been to show that novae can be  detected in typical MUSE observations of early-type galaxies out to $\sim$20\,Mpc. 

Novae have been observed in such galaxies many times in the past, and H$\alpha$-variability selection is a standard method for identifying them. To my knowledge, however, the current work is the first to detect novae at such distances in IFS data. 
The spectroscopic approach  can provide more optimised tuning of the selection bandwidth, as well as improved continuum subtraction. Moreover, of course, the IFS method yields the full spectrum simultaneously. 

Although the novae can be discovered in single observations, I have used multi-epoch data to help confirm their identification where available. Selecting {\it explicitly} for broad-H$\alpha$ variability would enable extension to galaxies such as M\,87, which was excluded here due to 
confusion from its complex nebular H$\alpha$ emission.
A possible future MUSE monitoring campaign observing with $\sim$1-month cadence would be able to detect and follow $\ga$20 novae per target galaxy per year, while simultaneously building a very deep `static' datacube that would find other applications, e.g. for detailed stellar populations and kinematic studies.

\section*{Acknowledgements} 
I thank Jay Strader for Twitter discussions which helped convince me that the sources were probably novae.  This work was supported by the Science and Technology Facilities Council through the Durham Astronomy Consolidated Grant 2017--2020 (ST/P000541/1).

\bibliographystyle{mnras}
\bibliography{rjs}

\bsp	
\label{lastpage}
\end{document}